\documentstyle[12pt,epsf,psfig]{article}

\newcommand{\be}{\begin{equation}}
\newcommand{\ee}{\end{equation}}
\def\reff#1{(\ref{#1})}

\begin{document}

\title{{\vspace{-0cm} \normalsize
\hfill \parbox{40mm}{CERN-TH/97-321}}\\[40mm]
Non-Perturbatively Improved \\ Quenched Hadron Spectroscopy}

\author{A. Cucchieri$^a$, M. Masetti$^a$, T. Mendes$^a$, 
         R. Petronzio$^{a,b}$ \\[1mm]
        {\small$^a$\em Dipartimento di Fisica, 
                 Universit\`a di Roma ``Tor Vergata''} \\[-1mm]
           {\small \em  and INFN, Sezione di Roma II } \\[-1mm]
        {\small \em Via della Ricerca Scientifica 1, 00133 Rome, Italy}\\
        {\small$^b$\em CERN, Theory Division, CH-1211 Geneva 23,
          Switzerland}
        }

\date{}
\maketitle

\begin{abstract}
We make a quenched lattice simulation 
of hadron spectroscopy at $\beta=6.2$
with the Wilson action non-perturbatively improved.
 With respect to the unimproved case, the estimate of the 
lattice spacing is less influenced by the choice
of input hadron masses.
We study also the effects of using an improved quark mass in the
fits to the dependence of hadron masses upon quark masses.
\end{abstract}

\vfill
\begin{flushleft}
\begin{minipage}[t]{5. cm}
  { CERN-TH/97-321}\\
November 1997
\end{minipage}
\end{flushleft}

\thispagestyle{empty}
\clearpage

\section{Introduction}
The computational cost of the extrapolation to the continuum limit of
lattice QCD simulations can be significantly reduced by using
improved actions, where the leading cutoff effects are cancelled by
suitable counterterms. It has been shown that O(a) on shell improvement
is achieved by adding to the usual Wilson action the clover term, 
with a coefficient that has been recently determined non-perturbatively
by the ALPHA collaboration \cite{Luscher,Luscher2}.

We present here the results of our study of hadron spectroscopy
using the nonperturbatively clover-improved Wilson action
\be
S \;=\, S_W \,+\,  c_{SW}\,a^5\,\frac{i}{4}
\sum_x {\overline \Psi}(x)\,\sigma_{\mu\nu}{\hat F}_{\mu\nu}\Psi(x)
\;\mbox{,}
\ee
where $S_W$ is the standard Wilson action and the second term
is the naive continuum limit of the clover term (see \cite{Luscher} 
for details). Preliminary results have been reported in \cite{proceedings}.

We consider a lattice volume of $24^3\times 48$ and a coupling $\beta = 6.2$.
According to \cite{Luscher,Luscher2}, we thus take $c_{SW}= 1.61375065 $. 
We choose the following values for the hopping 
parameter $\kappa$:
0.1240, 0.1275, 0.1310, 0.1340, 0.1345, 0.1350, 0.1352.
The simulations were carried out on the 512 processor
computer of the APE100 series at Tor Vergata.

Our statistics come from 104 quenched gauge configurations,
generated by a hybrid over-relaxation algorithm,
with each update corresponding to a heat-bath sweep followed by three 
over-relaxation sweeps. 
The configurations are separated by 1000 updates.

The inversion of the fermion matrix is performed using the 
stabilized biconjugate gradient algorithm \cite{BiCGStab}.
We restart the inversion from the current solution every 100 iterations,
in order to reduce the accumulation of roundoff errors \cite{refresh}.
We employ point-like sources.
We sum fermion propagators over the space directions $x,y,z$ for
sites within blocks of side 3, and then store the result. We then
form hadron correlations from these ``packed'' propagators. This procedure
differs from the exact computation of hadron correlations
by gauge-non-invariant terms and becomes exact in the limit 
of an infinite number of configurations (in our case we have checked that 
the errors thus introduced are negligible).
This corresponds to gaining a factor $3^3$ in storage, 
and has enabled us to have all quark propagators stored simultaneously. 
In this way we can form ``off-line'' hadron correlations from 
non-degenerate combinations of quark flavours.

Hadron masses are obtained from single mass fits to the
large time behaviour of zero momentum hadron correlators.
A two mass fit has also been done, but the results for
the higher mass were too unstable to quote numbers. 
The value for the lower mass turned out to be totally compatible with the
single mass fit.
The errors are estimated through a jack-knife procedure.
In order to improve the stability in time of the plateau where a single
hadron dominates the correlation for the baryons, in some cases  
we have averaged the correlation over a fixed number of lattice 
spacings in the time direction: the resulting correlation 
maintains its exponential behaviour with a coefficient depending upon
the size of the smearing. This procedure turned out to be very useful
for the determination of baryon mass splittings where we also made
fits directly to the ratio of correlations to minimize the effects
of collective fluctuations.
We report in Table \ref{tab:lat_data} our results for hadron masses 
in lattice units and at the various values of $\kappa$.

\begin{table}[htb]
\addtolength{\tabcolsep}{0mm}
\begin{center}
\begin{tabular}{cccc}
\hline
$\kappa $ & ${M_{PS}}$ & $M_V$  & $M_N $\\
\hline
0.1240 & 1.0737(20) & 1.0997(15)  & 1.7002(35)  \\ \hline
0.1275 & 0.8532(20) & 0.8873(15)  & 1.3774(35)  \\ \hline
0.1310 & 0.6048(20) & 0.6565(20)  & 1.0189(40)  \\ \hline
0.1340 & 0.3445(20) & 0.4364(35)  & 0.6608(65)  \\ \hline
0.1345 & 0.2909(25) & 0.3977(50)  & 0.5937(85)  \\ \hline
0.1350 & 0.2294(30) & 0.364(10)   & 0.520(15) \\ \hline
0.1352 & 0.2007(40) & 0.353(15)   & 0.483(20) \\ \hline
\end{tabular}
\caption{\label{tab:lat_data}
Masses in lattice units (diagonal-flavour combinations only)
for the pseudoscalar and vector mesons, and for the nucleon. }
\end{center}
\end{table}
\normalsize

\section{Lattice Spacing and Meson Masses}

In order to extract physical values from lattice data, one 
needs to interpolate the results at appropriate values of quark masses,
and to give them physical values by a suitable normalization.
These steps imply a number of choices.

The first concerns the determination of ``$\kappa_c$'': indeed,
the renormalization of the quark mass with Wilson fermions is not 
multiplicative, and the critical value of the hopping parameter is shifted 
from its free-case limit.
Within the improvement programme one can determine the critical value
$\kappa_c$ using the mass extracted from an improved Ward identity,
defined  as \cite{Luscher}
$$ m_{WI} \; \equiv \;
\frac{ \langle \partial_{\mu} \{ {A_{\mu}^{(bare)}} \,+\,
c_A \, a \, \partial_{\mu}{P^{(bare)}}\}{\cal O} \rangle}
{2 \langle{ P^{(bare)}}{\cal O} \rangle}
\;\mbox{.}
$$
with the parameter $c_A$ fixed from ref.\ \cite{Luscher} to $-0.037$.
A linear extrapolation to the limit of $\,m_{WI}=0\,$,using the first
five points obtained from the combinations 
of the three highest $\kappa$ values, provides a fit 
for the determination of $\kappa_c$ much more stable than the conventional 
fit of pseudoscalar meson masses to the limit of zero pion mass.
Our results for $\kappa_c$ are:
\begin{eqnarray}
\mbox{from} \quad m_{WI} = 0 \; & & \kappa_c = 0.135828(5) \nonumber \\[1mm]
\mbox{from} \quad M_{PS}^2 = 0 \; & & \kappa_c = 0.135849(13) .\nonumber 
\end{eqnarray}
The values of the parameters of a linear fit in $1/\kappa$ to
$M_{PS}^2$ and to $m_{WI}$ are:
\begin{eqnarray}
m_{WI} &=& -3.8059(15) \,+\, 0.5169(2) \times (1/\kappa) \\[1mm]
M_{PS}^2 &=& -8.38(3) \,+\, 1.138(4) \times (1/\kappa) . 
\end{eqnarray}
The residual discrepancy between the two values of  $\kappa_c$ is a signal of
residual lattice artefacts.

Once the choice for the value for $\kappa_c$ from $m_{WI}$
has been made, we make a second choice that improves
 the fits to the dependence of hadron masses
upon quark masses: instead of using the bare
quark mass, defined as 
$ m_q(\kappa) \,\equiv\, (1/\kappa\,-\,1/{\kappa_c})/2$,
we use  an ``improved'' bare quark mass \cite{Luscher} defined by
\be
{\widetilde m_q}(\kappa) \;\equiv\; m_q(\kappa)\,[1\,+\,b_m\,m_q(\kappa)]
\;\mbox{.}
\label{eq:mR}
\ee
(Note that ${\widetilde m_q}$ is the renormalized mass with $\,Z_m=1\,$.)
The improvement coefficient $b_m$ has been determined non-perturbatively 
\cite{Giulia}:
$ \,b_m = -0.62(3)\,$.

For non-degenerate flavour cases we use symmetric averages
of the masses defined above, e.g. for a meson corresponding
to the flavours $\kappa_1$ and $\kappa_2$, we define
$  {\widetilde m_q}(\kappa_1,\kappa_2) \;\equiv\; 
[{\widetilde m_q}(\kappa_1)\,+\, {\widetilde m_q}(\kappa_2)]/2$. 
Similarly, for baryons,
we define $m_q$ and ${\widetilde m_q}$ as the symmetric average
of the quark masses for the three flavours.
Note that when using averages of the improved masses we reabsorb
in the mass definition terms that are quadratic in the quark masses.
This may and does change the estimates of the deviation of hadron
masses (their squares for pseudoscalars) from a linear dependence upon
the quark masses. For baryons the choice of the improved mass does
 extend the linear behavior substantially beyond the light-mass region.\\

A third choice to be made is on the physical inputs that give physical units
to the lattice spacing and to the lattice masses.

The lightest of our quark masses is in the strange quark mass region and 
we decided to use physical inputs at the strange quark mass for the
lattice spacing to
avoid the inclusion of systematic uncertainties deriving from a
chiral extrapolation.
A known deficiency of the quenched approximation is the estimate of
vector-pseudoscalar mass splittings. In particular the experimental 
difference of the squares of vector and pseudoscalar masses is essentially 
constant as a function of the quark mass, a feature that is not 
reproduced by present quenched data.
Besides possible unquenching effects, this can be ascribed to lattice 
artefacts, which should be partially cancelled in our simulation.

For the strange mesons, we decided to use as input just such a square-mass 
difference:
this can be seen as an attempt to reabsorb the residual lattice artefacts
in a redefinition of the quark mass.
More in detail, we first fix the value of the strange quark mass from
the ratio:
\begin{equation}
(M_V^2 - M_{PS}^2)/M_V^2 
\end{equation}
using the experimental value for $K$ and $K^{*}$ meson masses
as input, and then we extract the lattice spacing by normalizing 
the $K^{*}$ mass to its experimental value. This gives the following values:
\begin{equation}
m_s \,=\, 0.0315(45) \quad \mbox{and} 
\quad a^{-1} \,=\, 2561(100) \;\;\mbox{MeV}
\;,
\label{eq:ms}
\end{equation}
where we have used, for the average values, quadratic and linear
mass fits, respectively.
These are the values of $m_s$ and $a^{-1}$ adopted in this paper.
In the following we will not include the overall error coming from 
the uncertainty of the lattice spacing when quoting errors for
hadron masses in physical units. This amounts to an overall and systematic
uncertainty of about 4\%.

Our fits for $M_V$ and $M_{PS}^2$ in the light/strange quark
mass region are given below.
\begin{itemize}
\item {\bf Linear} fits in terms of the {\bf improved} quark mass:
\begin{eqnarray}
M_{V} &=& 0.309(20) \,+\, 2.49(55) \times {\widetilde m_q} \\[1mm]
M_{PS}^2 &=& -0.001(2) \,+\, 2.398(35) \times {\widetilde m_q}
\end{eqnarray}
\item {\bf Quadratic} fits in terms of the {\bf improved} quark mass:
\begin{eqnarray}
M_{V} &=& 0.298(15) \,+\, 2.82(50) \times {\widetilde m_q}  
        \,+\, 1.0(35) \times {\widetilde m_q}^2 \\[1mm]
M_{PS}^2 &=& 0.005(2) \,+\, 1.992(50) \times {\widetilde m_q}
        \,+\, 7.53(40) \times {\widetilde m_q}^2
\end{eqnarray}
\item {\bf Linear} fits in terms of the {\bf unimproved} quark mass:
\begin{eqnarray}
M_{V} &=& 0.311(20) \,+\, 2.38(55) \times m_q \\[1mm]
M_{PS}^2 &=& 0.000(2) \,+\, 2.328(30) \times m_q
\end{eqnarray}
\item {\bf Quadratic} fits in terms of the {\bf unimproved} quark mass:
\begin{eqnarray}
M_{V} &=& 0.304(15) \,+\, 2.60(45) \times m_q  
         \,+\, 0.4(30) \times {m_q}^2 \\[1mm]
M_{PS}^2 &=& 0.004(2) \,+\, 2.085(50) \times m_q
         \,+\, 3.91(35) \times {m_q}^2
\end{eqnarray}
\end{itemize}

The quadratic fit for the pseudoscalar meson gives evidence for a
quadratic term, while for the vector case its presence appears very marginal.
The  coefficient of the quadratic term in the unimproved mass case is expected to be
smaller than in the improved case from the relation in eq.\ \reff{eq:mR}.

The chiral limit of vector masses is obtained by linear extrapolation
in terms of the improved quark mass, showing good agreement
when compared to the experimental value. This indicates that a lattice 
spacing extracted from the chiral limit leads to a value very close 
to the one in eq.\ \reff{eq:ms}:
\begin{equation}
 a^{-1}_{chiral} \;=\; 2486(175) \;\;\mbox{MeV.}
\end{equation}

The fit for the vector case goes through the $\phi$ mass with reasonable
accuracy and indicates that in the strange quark mass region a lattice spacing 
normalization from the $\phi$ would also give similar results:
\begin{equation}
 a^{-1}_{\phi} \;=\; 2634(40) \;\;\mbox{MeV.}
\end{equation}

\begin{figure}
\centerline{\psfig{figure=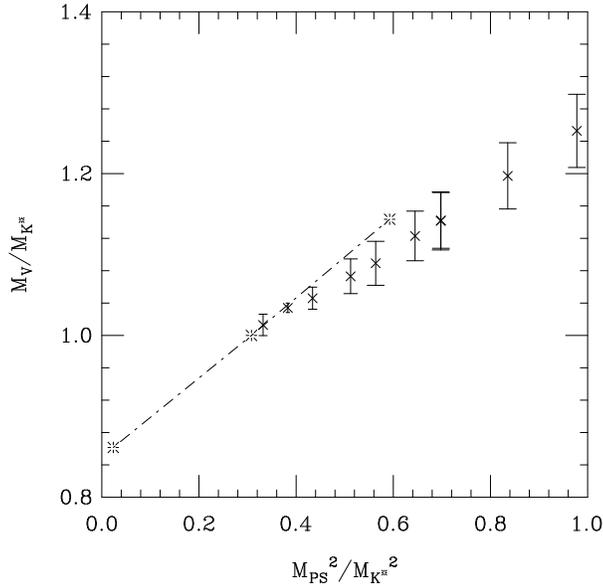,height=3.5in}}
\caption{APE plot for $M_V$.}
\label{fig:APE_rho}
\end{figure}

In Fig.\ \ref{fig:APE_rho} we show a plot of the vector meson
mass as a function of $M_{PS}^2$.
The masses are normalized to the $K^{*}$ meson mass, taken at the strange 
quark mass value given in the eq.\ \reff{eq:ms} above. The asterisks
correspond to the physical values for the ratio.

As can be seen from this figure, the experimental slope $\,dM_V/dM_{PS}^2\,$
is different from the one corresponding to our data. This is related
to the behaviour of the quantity $J$ defined by:
\begin{equation}
J \;\equiv\; M_V\,(dM_V/dM_{PS}^2)
\;,
\end{equation}
obtained from $M_V$ as a function of $M_{PS}^2$.
Our results for $J$ are quoted in Table \ref{tab:J}.
They are rather similar to those obtained without improvement 
(for a recent review, see \cite{J_unimproved}), although 
slightly closer to the experimental result.
 
\begin{table}[htb]
\addtolength{\tabcolsep}{1mm}
\begin{center}
\begin{tabular}{ccc}
\hline
                & Exp. value & Our value \\
\hline
$ J_{K^{*}}$ & 0.487 & 0.40(11)  \\
\hline
$ J_{\phi} $ & 0.557 & 0.45(12)   \\
\hline
\end{tabular}
\caption{\label{tab:J}
Values of the quantity $J$ and comparison with experiment. 
}
\end{center}
\end{table}
\normalsize

\vskip 3mm
The procedure used to determine the strange quark mass fails when
applied to the charm case: it is not possible here to find a value
of the quark mass that reproduces the observed splitting.
Therefore, we decided to define the charm quark mass from a fit
to the ratio $M_{PS}^2/M_{K^{*}}^2$.
We get
\begin{eqnarray}
m_c &=& 0.388(25)  \;\quad \mbox{from} \;\;  m_q ,   \\[1mm]
m_c &=& 0.3154(10)  \;\quad \mbox{from} \;\;  {\widetilde m_q} .
\label{eq:mc}
\end{eqnarray}

By using the ``$K^{*}$'' 
lattice spacing we obtain for the $D^{*}$ charmed meson: 
\begin{equation}
M_{D^{*}} \;=\; 1955(6) \;\;\mbox{MeV}  
     \quad \mbox{(exp. value 2008 MeV)} .
\end{equation}
The above value for the vector mass comes from a linear fit to a cluster
of points around the improved charm quark mass.

For the vector meson with an unimproved quark mass we find:
\begin{equation}
M_{D^{*}} \;=\; 1981(5) \;\;\mbox{MeV} . 
\end{equation}
In both cases, the underestimate of 
the experimental pseudoscalar-vector mass splitting is not dramatic.

The value of the lattice-improved strange quark mass given in
eq.\ \reff{eq:ms} can be used to obtain the strange quark mass in
physical units. Firstly, we multiply it by the mass renormalization 
factor $Z_m$ at the $(1/a)$ scale, using the tadpole-improved
formula given in eq.\ (41) of ref.\ \cite{Schierholz}. We then evolve the
result to the 2 GeV scale with a perturbative renormalization group 
factor as in \cite{Gupta}. We get
\begin{equation}
m_s \;=\; 111(15) \;\mbox{MeV}\;.
\label{eq:ms_phys}
\end{equation}
This result is in agreement with that of ref.\ \cite{Schierholz}.

In the ratio of charm over strange quark mass the renormalization
factor $Z_m$ 
drops out and we can give a value for the ratio of renormalized quark masses:
\begin{eqnarray}
m_c/m_s &=& 12.1 \pm 1.6 \;\quad \mbox{from} \;\; m_q , \nonumber \\[1mm]
m_c/m_s &=& 10.0 \pm 1.4 \;\quad \mbox{from} \;\; {\widetilde m_q} .
\nonumber
\end{eqnarray}
Notice that the difference comes essentially from the reduction of the
charm quark mass going from the unimproved to the
improved bare quark mass definition (see eq. \reff{eq:mc}).

We can use this ratio, combined with the theoretical prediction of
1.220(60) GeV for the charm quark mass \cite{mc_theo}, to obtain an
independent evaluation of the strange quark mass. After evolving this
value of the charm quark mass to the 2 GeV scale, we get:
\begin{eqnarray}
m_s &=& 92(13)\;\mbox{MeV} \;\quad \mbox{from} \;\; m_q ,
             \nonumber \\[1mm]
m_s &=& 111(16) \;\mbox{MeV} \;\quad \mbox{from} \;\; {\widetilde m_q} .
             \nonumber
\end{eqnarray}
The value obtained using the improved bare quark mass is in very good
agreement with the determination in eq.\ \reff{eq:ms_phys}.

\section{Baryon Masses}
In this section we report our results for baryon masses and baryon
mass splittings. We show APE plots for the octet and for the
decuplet baryons in Fig.\ \ref{fig:APE}.
Note that we divide at each point by $M_V$,
interpolated to the strange-quark mass, corresponding to $M_{K^{*}}$
(i.e. a constant value). Experimental points in these figures (asterisks)
correspond to $M_N$, $M_{\Sigma}$ and $M_{\Xi}$
and appropriate meson masses in Fig.\ \ref{fig:APE}a, and similarly
$M_{\Delta}$, $M_{\Sigma^{*}}$,$M_{\Xi^{*}}$ and $M_{\Omega}$ for
Fig.\ \ref{fig:APE}b.

\begin{figure}
\centerline{\psfig{figure=pnr_APE_cswnp.ps,height=3.5in}}
\centerline{\psfig{figure=dlt_APE_cswnp.ps,height=3.5in}}
\caption{~APE plot for (a) the nucleon mass (octet baryons) and
(b) the $\Delta$ mass (decuplet baryons).}
\label{fig:APE}
\end{figure}

We have considered non-relativistic wave functions.
The correlators for the spin-3/2 (decuplet) baryons are
completely symmetric in flavour. We therefore expect
to observe a smooth behaviour of their masses in terms of
the average quark masses $m_q$ and ${\widetilde m_q}$.
For the spin-1/2 (octet) baryons, on the contrary, the correlators are not
completely flavour-symmetric. We have two types of correlators (see
\cite{Sharpe} for details): $\Sigma$-like (e.g. for the
proton, neutron, $\Sigma$ and $\Xi$) and $\Lambda$-like 
(e.g. for $\Lambda$). The expressions for their masses from
quenched chiral perturbation theory, in the case of two light quarks 
$m_u$ and a strange quark $m_s$, are given
by \cite{Sharpe}:
\begin{eqnarray}
M_{\Sigma} &=& M_0 \,+\, 4\,F\,m_u \,+\, 2\,(F - D)\,m_s \\[1mm]
M_{\Lambda} &=& M_0 \,+\, 4\,(F - 2\,D/3)\,m_u
    \,+\, 2\,(F + D/3)\,m_s .
\end{eqnarray}
The constant $D$ can be related to the $\Sigma$--$\Lambda$ octet
hyperfine splitting, as can be seen from the above formulae.
We have considered fits as functions of the symmetric average
masses described above also in the spin-1/2 case.

Note that we have defined the quantity
\begin{equation}
M_{oct} \equiv (M_N + M_{\Lambda})/2
\;,
\end{equation}
which we construct from the appropriate ratios of correlators.
This combination was chosen so that the resulting mass is
flavour-symmetric. For this ``octet'' particle the advantage of using the
improved quark mass is clearest, see Fig.\ \ref{fig:oct}.

From fits of the $\Sigma$--$\Lambda$ mass splitting, and of $M_{oct}$,
it is possible to extract the value of the constants $D$ and $F$. 
We obtain:
\begin{eqnarray}
D &=& -0.51(20) \\
F &=& 0.80(20).
\end{eqnarray}

\begin{figure}
\centerline{\psfig{figure=oct_mpert.ps,height=3.5in}}
\centerline{\psfig{figure=oct_mR.ps,height=3.5in}}
\caption{~``Octet'' mass in terms of (a) $m_q$ (b) ${\widetilde m_q}$.}
\label{fig:oct}
\end{figure}

In Table \ref{tab:masses1} we give our baryon mass values in MeV.
We have used in all cases linear fits
of the masses as functions of the average improved bare quark mass 
${\widetilde m_q}$,
except for the $\Sigma$--$\Lambda$ mass splitting, where
we used a quadratic fit. We have also included the predictions for charmed
baryons using the improved quark mass.

%
\begin{table}[htb]
\addtolength{\tabcolsep}{0mm}
\begin{center}
\vspace{-0.5cm}
\begin{tabular}{cccc}
\hline
   & Exp. value & Our value \\
\hline
$ M_{N} $                  & $ 939      $ & $ 952(110) $ \\
\hline
$ M_{\Sigma} $             & $ 1189.4   $ & $ 1148(60) $  \\
\hline
$ M_{\Sigma-\Lambda} $     & $ 73.7     $ & $ 70(30) $ \\
\hline
$ M_{\Delta} $             & $ 1232     $ & $ 1265(110) $ \\
\hline
$ M_{\Delta-N} $           & $ 293      $ & $ 297(80) $ \\
\hline
$ M_{\Sigma_c} $           & $ 2453     $ & $ 2406(15) $  \\
\hline
$ M_{\Sigma_c-\Lambda_c} $ & $ 168      $ & $ 232(40)  $  \\
\hline
\end{tabular}
\caption{\label{tab:masses1}
Baryon masses in MeV and comparison with experiment. The error in 
parentheses represents the statistical error, while the error due
to the determination of the inverse lattice spacing is of about 4\%.}
\end{center}
\end{table}
\normalsize

\section{Conclusions}
The main effect of the
non-perturbative improvement observed on hadron masses is
a smaller spread in the value of the lattice spacing extracted
from light mesons with and without strange quarks.
We have also shown how the use of the improved quark mass 
turns the rough behaviour of the dependence of the octet baryon 
mass upon quark masses into a smooth one.
We have found a remarkable agreement in the improved theory
between  two independent determinations of the strange quark
mass, one normalized through the lattice spacing and the other from the
value for the charm mass  extracted in the continuum from charmonium spectrum
calculations.

For an accurate comparison of these results with the unimproved
case, we are presently analysing 
\cite{csw0} on the same set of gauge configurations the case $c_{sw} = 0$.

\end{document}